\documentstyle[12pt]{article}
\newlength{\abstractwidth}
\def \be {\begin{equation}}
\def \ee {\end{equation}}
\def \bea {\begin{eqnarray}}
\def \eea {\end{eqnarray}}

\begin{document}
\bigskip
\bigskip\bigskip\bigskip\bigskip
\bigskip \bigskip
\centerline{\Large \bf {Note on Holographic RG Flow in String
Cosmology}}
\bigskip\bigskip
\bigskip\bigskip
\centerline{Miao Li$^{1}$  and Feng-Li Lin$^{2}$}
\bigskip
\centerline{\it ${}^1$Institute of Theoretical Physics}
\centerline{\it Academia Sinica} \centerline{\it Beijing 100080}
\centerline { and} \centerline{\it Department of Physics}
\centerline{\it National Taiwan University} \centerline{\it Taipei
106, Taiwan} \centerline{mli@phys.ntu.edu.tw}
\bigskip
\centerline{\it ${}^2$Department of Physics} \centerline{\it
Tamkang University} \centerline{\it Tamsui, Taipei 25137, Taiwan}
\centerline{linfl@mail.tku.edu.tw}
\bigskip\bigskip

%ABSTRACT

\begin{abstract}
We propose a new holographic C-function for the accelerating
universe defined in the stringy frame motivated mainly by the fact
that the number of degrees of freedom should be infinite for a
physical system of infinite size. This is the generalization of
Strominger's recent proposal of the holographic C-function to the
asymptotically non-de Sitter universe. We find that the
corresponding C-theorem holds true if the universe accelerates
toward the weak coupling regime driven by the exponential dilaton
potential. It also holds in other simple cases.

\end{abstract}
%\pacs{PACS numbers: \ }

\newpage

\section{Introduction}

Recent results in observational cosmology provide substantial
evidence for the existence of dark energy \cite{nova}. String
theory for the first time has to cope with a possible positive
cosmological constant, or a tiny positive vacuum energy which may
depend on some moduli. As pointed out in \cite{Q}, with either a
cosmological constant or the popular quintessence model, the
universe's expansion will forever accelerate, resulting in a
future horizon. This poses conceptual problems in string theory
thus far formulated \cite{Witten}.

Despite the fact that no one has come up with a credible model
with a positive vacuum energy in string theory (for some attempts,
see \cite{Hull}), people have speculated on the possible
microscopic descriptions of such a universe.  This includes a
possible dS/CFT correspondence which generalizes the usual AdS/CFT
correspondence, and a possible holographic horizon theory and
matrix models \cite{mlg}. The dS/CFT approach has since been
followed up by many authors \cite{dscft}. It appears to us that
both lines of approach contain some ingredients of truth and may
even be related to each other. Both are partially motivated by the
conjecture that the number of degrees of freedom in a de Sitter
universe is finite \cite{banks}. It is thus a good question to ask
whether at a given time, the number of degrees of freedom
observable to a given observer is finite and can be written as a
local function of geometry at that given time. By analogy with the
AdS/CFT correspondence, there is an answer if the asymptotic
geometry in the future is de Sitter, as discussed in
\cite{strominger2} and the second reference of \cite{strominger1}.

The central charge, or more generally, the measure of the number
of degrees of freedom defined in \cite{strominger2} for a geometry
asymptotating de Sitter space is given by
\be
C(t)={1\over H^{d-1}G},
\label{cc}
\ee
where $H$ is the Hubble constant and $G$ is the Newton constant in
$d+1$ dimensions. For a de Sitter space, $H=1/R$, $R$ is the
radius of the cosmic horizon, the formula (\ref{cc}) coincides
with the one for the cosmic entropy. For a geometry asymptotically
de Sitter, $C(t)$ is a function of time, and the formula can be
inferred by comparing to the known formula in the AdS/CFT
correspondence \cite{gubser}. One must keep in mind that such
formula can be related to the Weyl anomaly only when the boundary
theory lives in an even dimension, that is, when $d$ is even. For
odd $d$, in particular for $d=3$ of the observed universe,
(\ref{cc}) stands as a good guess. Now just as in the AdS/CFT
correspondence, Einstein equations together with the null energy
condition ensure nondecrease of the central charge function.

When one claims that the central charge increases as the universe
evolves, one is not claiming a law similar to the second law of
thermodynamics, since in calculating ${dC\over dt}$ one has not
resorted to thermodynamics at all. Actually an arrow of time is
already chosen by assuming the positivity of the Hubble constant
$H$. In an odd dimensional universe, $H$ has to be positive for
$C$ to be positive. In an even dimensional universe, $H$ does not
have to be positive in defining $C$, however in showing ${dC\over
dt}\ge 0$, one has to assume $H\ge 0$.

While (\ref{cc}) may well be a good candidate for the C-function
in a universe asymptotically de Sitter, it may fail for more
general situation when the fate of the universe is completely
different. For instance, in a universe driven by quintessence in
the future, the size of the horizon increases indefinitely, one
does not expect the same formula to hold here. This is already the
case in the AdS/CFT correspondence. One example is the gravity
dual of noncommutative Super Yang-Mills, where the asymptotic
geometry is not AdS, and naturally the C-function must be worked
out separately as in \cite{lw}. In the present context, one may
consider the case when the acceleration of the expansion vanishes,
the borderline of the quintessence model. The radius factor $a(t)$
is proportional to $t$, so the physical size of the future horizon
\be
R(t)=a(t)\int_t^\infty {dt\over a(t)},
\label{hsize}
\ee
diverges for any finite time $t$, thus one expects the measure of
the number of degrees of freedom also diverge. Formula (\ref{cc})
however gives a finite number.

If the quintessence field is the dilaton, we shall see that the
Hubble constant in the stringy frame actually vanishes if the
string coupling is driven to the weak coupling regime, this
motivates us to propose the definition
\be
C(t)={1\over H_s^{d-1}G_s}.
\label{scc}
\ee
Now the Newton constant in the stringy frame depends on the
dilaton thus depends on time in general. The above formula is
taken to be a good candidate only in the weakly coupled string
theory. We shall show that for known quintessence potentials, a
C-theorem is valid for the above definition. Actually, a C-theorem
holds only when the universe expansion accelerates if the
potential is an exponential of the dilaton. Thus, C-theorem
formulated in terms of (\ref{scc}) is more restrictive than the
one formulated in terms of (\ref{cc}). We take this as a good
sign, namely not all solutions to the Einstein equations with a
reasonable matter content are all plausible universes.

The strongest support to the formula (\ref{scc}) comes from the
following fact. For an exponential quintessence potential for the
dilaton $\phi$, the solution $a(t)$ scale as $t^{1+\kappa}$. In
this case, the future horizon size (\ref{hsize}) is $(1/\kappa)
t$. As $\kappa \rightarrow 0$, this blows up. The area of the
future horizon then scales as $(1/\kappa^{d-1}) t^{d-1}$. It
happens that in this case the central charged defined in
(\ref{scc}) also scale in the same way in terms of both $\kappa$
and $t$.

If the universe is driven to the strongly coupled regime far into
the future, one may either use (\ref{scc}) or a different
definition motivated by M theory. If (\ref{scc}) is used, the
C-theorem always holds. If a M theory definition is adopted, we
shall see that a corresponding C-theorem also holds, and the
condition is simply $p<\rho$. However, no natural definition will
give an infinite $C$ for the case $a(t)\sim t$. Does this
impossibility imply that this case is impossible in the M theory
regime, or alternatively the fate of the string coupling is zero
in the far future?

Our discussions in the following will be focused in the far
future, since there one is devoid of the problem of complicated
matter as the dark energy dominates.

 \section{Holographic C-theorem of the FRW Universe in the Einstein
frame} For more general consideration, let us start with the
action of the dilatonic gravity in d+1+D dimensions for string
cosmology \cite{stringcosmology}:
\be
S=S_0+S_U=-\int d^{d+1+D}X \;\sqrt{-G} e^{-2\phi}[R+4 (\nabla
\phi)^2 -U(\phi)]\;,
\label{full}
\ee
where $U(\phi)$ is the dilaton potential which is zero at tree
level for critical string(if $d+1+D=d_c$, the critical dimension)
but can get nontrivial quantum corrections; and for the
noncritical string $U={2\over3}(d_c-d-1-D)$ at tree level.

Compactifying the theory on D-dimensional torus $T^D$ with the
following metric ansatz
\be
ds^2=G_{MN}dX^M dX^N=\hat{g}_{\mu \nu}(x)
dx^{\mu}dx^{\nu}+\sum^{D}_{i=1} e^{2\sigma_i{(x)}}dy_i^2\;,
\ee
where $\hat{g}_{\mu \nu}$ is the metric of the (d+1)-dimensional
noncompact space. After some manipulations, we arrive the
following dimensionally-reduced action\footnote{Set the volume of
$T^D$ parameterized by $y_i$ equal to one.}:
\be
\label{stringya}
S=-\int d^{d+1}x \;\sqrt{-\hat{g}}
e^{-\Phi}[\hat{R}+(\hat{\nabla}\Phi)^2-\sum^D_{i=1}(\hat{\nabla}
\sigma_i)^2-U(\Phi,\sigma_i)]\;,
\ee
where the hat quantities are with respect to metric $\hat{g}_{\mu
\nu}$, and the new field $\Phi\equiv
2\phi-\sum^{D}_{i=1}\sigma_i$. Also note that the moduli
$\sigma_i$ could get a nontrivial potential from quantum
correction.

We transform the action into the Einstein one by
\be
g_{\mu\nu}=e^{-2\Phi \over (d-1)} \hat{g}_{\mu\nu}\;,
\label{Weyl}
\ee
and the action becomes
\be
S=-\int d^{d+1}x \;\sqrt{-g} [R-{1\over d-1} (\nabla
\Phi)^2-\sum^D_{i=1}(\nabla \sigma_i)^2-V(\Phi,\sigma_i)]\;,
\ee
where $V(\Phi,\sigma_i)\equiv e^{2\Phi \over d-1} U(\Phi,
\sigma_i)$.

    From this relation, a potential $U=\sum_{n=1}c_n g^{2(n-1)}$
calculated from perturbative string theory will be transformed
into $g^{4\over d-1}U$ in the Einstein frame, where $n$ is the
number of loop and the string coupling
\be
g=e^{\Phi/2}\;.
\ee
Moreover, the 1-loop potential $V=e^{{2\over d-1}\Phi}$ is
dominant as $\Phi\rightarrow -\infty$ but is just marginally able
to drive the universe in acceleration as remarked in
\cite{banksdine1}. In this note we assume that the expansion of
the Universe is driven by a generic potential $V$ in the late time
so that we need to require $V$ to be a slowly varying {\it
positive} function for our Universe to be de Sitter-like to
conform with the astronomical observation in \cite{nova}.

 Note that both the ``dilaton" $\Phi$ and the internal moduli $\sigma_i$
become canonical scalars in the Einstein frame, however, a
nontrivial potential mixes these scalars.  Moreover, from the
action we can derive the energy density and pressure if we treat
the "matter" part as perfect homogeneous fluid, it turns out to be
\bea
\label{energydensity}
\rho&=& {1\over d-1} \dot{\Phi}^2+\sum^D_{i=1} \dot{\sigma}_i^2+
V(\Phi,\sigma_i)\;,
\\
p&=&{1\over d-1}  \dot{\Phi}^2+\sum^D_{i=1}
\dot{\sigma}_i^2-V(\Phi,\sigma_i)\;,
\eea
where the dot is the derivative with respect to the time
coordinate $t$ of the FRW metric
\be
\label{FRW}
ds_{FRW}^2=g_{\mu\nu}dx^{\mu}dx^{\nu}=-dt^2+e^{2\lambda(t)}d\Sigma_k,
\ee
where $k=-1,0,1$ for open, flat and closed Universes.

The FRW equations are\footnote{Here we set $16\pi G_N$ equal to
one.}
\bea
\label{FRW1}
H^2&=&{-k\over a^2}+{1\over d(d-1)}\rho\;,
\\
\label{FRW2}
\dot{H}&=&{k\over a^2}-{1\over 2(d-1)}(p+\rho)\;,
\eea
and the field equations for the scalars are
\bea
{2\over d-1}(\ddot{\Phi}+dH\dot{\Phi})+{\partial V\over \partial
\Phi}&=&0\;,
\\
2(\ddot{\sigma}_i+dH\dot{\sigma}_i)+{\partial V \over \partial
\sigma_i}&=&0\;.
\eea
where we have defined the Hubble constant $H=\dot{\lambda}$
according to the scale factor $a=e^{\lambda(t)}$.

    From the fact of $p+\rho={2\dot{\Phi}^2 \over d-1}>0$ one can
immediately see that $H$ is monotonically decreasing for $k\le
0$\footnote{We will restrict ourselves to k=0 case only in the
following discussions.}.  This is the key observation of C-theorem
for the holographic dual CFT in \cite{strominger1} that the
central function (\ref{cc}) is monotonically increasing during
time evolution as long as $\rho$ tends to constant in the future
infinity such that the universe is asymptotically de Sitter, which
is a UV fixed in the dual CFT picture. Although this coincidence
strengthen the dS/CFT correspondence proposed in
\cite{strominger2}, it seems less powerful to constrain the
possible cosmological scenarios since it is valid for very generic
matters as long as the general positive theorem holds. This is in
contrast to the Fischler-Susskind cosmic holographic conjecture
\cite{FS1} which gives constraints to the possible cosmological
behavior.

Another puzzle regarding the holographic C-theorem is that the
C-theorem is not time-reversal invariant but the field equations
are. How can the second order field equations know the 2nd law of
thermodynamics which is implied by the increase of the C-function
as the Universe evolves, or the number of degrees of freedom in
the dual CFT? The resolution to the puzzle is that we have
required the Hubble constant to be positive, which is odd under
time-reversal so that a time direction is picked up when we assume
the expanding Universe. Especially for even $d$, to ensure the
positivity of the C-function, the condition $H>0$ should be
required for the notion of C-function to be sensible.

\section{Accelerating Universe}
The cosmological behavior should not be frame-dependent and two
different frames are related by just field redefinitions, however,
for weakly coupled strings, strings are the more natural probes,
or fundamental objects, so the stringy frame is favorable.
Moreover, physics should be different in the different regime of
the coupling constant in accord with the cosmic holographic RG
picture, for strong string coupling the M-theory frame is more
suitable where the 11-th dimension opens up. It is then also
reasonable to consider the holographic dual picture in the stringy
frame for weak coupling and M-theory frame for strong coupling,
then the stringy nature may yield the new feature in the
holographic consideration. We will see that this is indeed the
case and the holographic C-theorem may serve as a constraint to
the possible cosmological scenarios.

Since the equations of motion alone are not enough to verify the
C-theorem in stringy frame as we shall see later, we need to
integrate out the equations of motion for specific checking. For
simplicity, we freeze all the moduli but one which is defined as
the length variable of the moduli space as $Z=\int dt
\sqrt{{1\over d-1} \dot{\Phi}^2+\sum^D_{i=1} \dot{\sigma}_i^2}$
\cite{banksdine1}, then $\rho=\dot{Z}^2+V(\Phi(Z),\sigma_i(Z))$
and the new equation of motion for $Z$ by combining the ones for
$\Phi$ and $\sigma_i$ is
\be
2(\ddot{Z}+dH\dot{Z})+{\partial V \over \partial Z}=0\;,
\ee
where we have used ${\partial V \over \partial Z}\dot{Z}={\partial
V \over
\partial \Phi}\dot{\Phi}+\sum^D_{i=1} {\partial V \over \partial
\sigma_i}\dot{\sigma_i}$.

Using the equations of motion by starting with
$\dot{\rho}=2\dot{Z}\ddot{Z}+{\partial V \over \partial Z}\dot{Z}$
and the "velocity" formula
\be
\dot{Z}=\pm \sqrt{\rho-V}\;,
\ee
we can arrive
\be
\rho =\mp 2\sqrt{d \over d-1}\int dZ \;\sqrt{\rho(\rho-V)}\;.
\label{rhoeqm}
\ee
The particle horizon and the scale factor can also be put in the
integral form as
\bea
R_H&=&\int {dt \over a}=\pm\int dZ {1\over a \sqrt{\rho-V}}\;,
\\
a&=&exp\{\pm{\sqrt{1\over d(d-1)} \int dZ \; \sqrt{{\rho \over
\rho -V}}}\;\}\;.
\eea
Qualitatively if energy density $\rho$ is asymptotically dominated
by the potential, then $a$ will diverge more rapidly than
$(\rho-V) \rightarrow 0$ in the late time such that the integral
defining $R_H$ is finite for all time, i.e. there exists the
future horizon which naturally arises in de Sitter space and the
accelerating Universe driven by quintessence as shown in \cite{Q}.
On the other hand, the accelerating Universe requires
\be
\label{acce1}
{\ddot{a} \over a}={1\over
d(d-1)}[-\dot{\Phi}^2-(d-1)\sum^D_{i=1}\dot{\sigma}_i^2+V(\Phi,\sigma_i)]>0\;,
\ee
which leads to a constraint on the Hubble constant
\be
\label{acce2}
H^2>{1\over d-1}[{\dot{\Phi}^2 \over
d-1}+\sum^D_{i=1}\dot{\sigma}_i^2]={1\over d-1}(\rho-V)\;,
\ee
where we have combined (\ref{FRW1}) and (\ref{acce1}) together to
arrive at (\ref{acce2}). Moreover, using (\ref{FRW1}) and the
relation $p=\rho-2V$, then (\ref{acce2}) is equivalent to the
constraint on the equation of state
\be
\label{prho}
\omega \equiv {p\over \rho}<{2-d \over d}\;,
\ee
which is also known as the bound for the quintessence matters.

Specifically we can choose a simplest form of
potential\footnote{For the moment we will freeze the moduli
$\sigma_i$ for all $i$, therefore $Z=\Phi/\sqrt{d-1}$.}, the
exponential potential $V \propto e^{\pm \alpha \Phi}$, with
$\alpha\ge 0$. As shown in \cite{Peebles} there exists an
attractive fixed point in the solution space such that $\omega$ is
constant in time. If we choose the branch with $V \propto
e^{-\alpha \Phi}$, we need to require that $\dot{\Phi}>0$ to agree
with the physical cosmology of decreasing not increasing dark
vacuum energy, and from (\ref{energydensity}) we have
\be
\label{strong}
\dot{\Phi}=\sqrt{(d-1)(\rho-V)}\;.
\ee
This leads to that the string coupling $g=e^{\Phi/2}$ evolves
toward the strong string coupling regime. We will refers to this
as the {\it strong coupling branch}. The other branch is to choose
$V \propto e^{\alpha \Phi}$ such that the energy density diminish
asymptotically, so does the string coupling; we refer to this as
the {\it weak coupling branch} with
\be
\label{weak}
\dot{\Phi}=-\sqrt{(d-1)(\rho-V)}\;.
\ee

    From (\ref{rhoeqm}) we can relate $\alpha$ to $\omega$ by
\be
\label{alphar}
\alpha = {\sqrt{2d(1+\omega)} \over d-1}
\ee
for both the strong and weak coupling branches. Using
(\ref{alphar}) the accelerating Universe condition (\ref{prho})
can be translated into constraint on the decaying rate of the
potential as
\be
\label{Q}
\alpha < {2\over d-1}
\ee
for the exponential potential case \cite{Peebles}.

   Another simple form of the quintessential potential is the power-law
one given by\footnote{To distinguish from the results for the case
of exponential potential we instead choose not $\Phi$ but the
length variable $Z$ as the only unfrozen moduli.}
\bea
\rho&=&\rho_0 (-Z)^{-\delta}\;,
\\
\rho-V&=&C_0 (-Z)^{-\gamma}\;,
\eea
where $a,b$ are positive constants and we have chosen the weak
coupling branch so that $\rho \rightarrow V$ as $Z\rightarrow
-\infty$ and
\be
\dot{Z}=-\sqrt{\rho-V}\;.
\ee
Note that in this case $\rho/V$ is no longer a constant so that it
leads to no fixed point in the solution space but to the tracker
solution \cite{steinhardt}.

   From (\ref{rhoeqm}) we can obtain the following relations
\bea
C_0&=&{(d-1)\delta^2\rho_0 \over 4d}\;,
\\
\gamma&=&\delta+2\;.
\eea
It is then easy to see that the Universe is accelerating for large
enough $Z$ for any $a$ and $\rho_0$ which instead will be
constrained by the condition of no additional long range force
besides the known ones.

\bigskip

\section{Holographic C-Theorem in Stringy frame}
As emphasized in the context of string cosmology, when strings are
weakly coupled it is natural to use the string probe in the
stringy frame to consider the cosmic holographic principle. In the
stringy frame the Hubble constant and the effective Newton
constant are different from the ones in Einstein frame and are
defined with respect to the stringy metric $\hat{g}_{\mu\nu}$ via
(\ref{Weyl}) and (\ref{FRW}), the results are
\bea
H_s&=&g^{{-2 \over d-1}}(H+{\dot{\Phi}\over d-1})\;,
\\
G^{(s)}_N&=&g^2 G_N\;,
\eea
where $G_N$ and $H$ are the Newton constant and Hubble constant
respectively in the Einstein frame, and $G_N$ is time-independent
by definition. From these, the inverse C-function is
\be
\label{C}
H_s^{d-1}G^{(s)}_N=G_N(H+{\dot{\Phi}\over d-1})^{d-1}\;.
\ee
  Note that although the scale behavior of $H_s$ is different from the
one of $H$ due to the dressed string coupling factor $g^{-2\over
d-1}$, the resultant C-function has the same scale behavior as the
one in the Einstein frame proposed by Strominger
\cite{strominger2} but differs by a sub-leading term. This fact is
essential for the C-theorem in the stringy frame to hold and be
closely related to the C-theorem in the Einstein frame. For the
special case of that $H$ and $\dot{\Phi}$ have the same scale
behavior, the resultant C-function differs from the one in the
Einstein frame only by an overall coefficient which is irrelevant
to the validity of the C-theorem as long as the coefficient is
positive.

For the string cosmology to make sense it requires that the
stringy Hubble constant is positive to agree with the expanding
universe observation even for an observer in the stringy frame.
This condition is also necessary for the holographic C-function in
the stringy frame to be sensible when $d$ is even. For the {\it
strong coupling branch} this is always true since $H>0$,
$\dot{\Phi}>0$. On the other hand, for the {\it weak coupling
branch} $\dot{\Phi}<0$,  the condition for $H_s>0$ means that
\be
\label{Hs}
H^2>{\dot{\Phi}^2 \over (d-1)^2}\;.
\ee
Surprisingly, this is coincident with the condition (\ref{acce2})
in the Einstein frame for the accelerating Universe driven by
dilaton alone but freezing other moduli\footnote{The moduli
dynamics is in fact positive to the constraint (\ref{Hs}).}. {\it
We conclude that the Universe in the weak string coupling branch
is observed to be expanding but not contracting if it is
accelerating in the Einstein frame.}  We have no clue of physical
reasons for this coincidence.

 The other reason to justify our choice of C-function (\ref{scc}) as
remarked in the Introduction is to look into the borderline case
of $H_s=0$. Its scale factor is linear in $t$, i.e. $a(t)=t$ so
that the size of the future horizon (\ref{hsize}) diverges but the
inverse Hubble constant in Einstein frame $1/H$ is finite, so is
the C-function (\ref{cc}). Although there is ambiguity about where
the holographic dual theory should locate, we believe that the
dual theory should live on a space with its size comparable to the
size of the future horizon. The finiteness of the C-function for
infinite future horizon is then incompatible with the conventional
wisdom of quantum theory that infinite physical system should
contain infinite number of degrees of freedom. Part of the reason
for the above failure of (\ref{cc}) is that the asymptotic
geometry of $H_s=0$ universe is not de Sitter. It is then
physically motivated to find the appropriate C-function for the
asymptotically non-de Sitter universes, which are generic for the
string cosmology with dilaton as the dominant driving force of the
cosmological expansion in the late time. Since the driving force
of the expansion of the universe is assumed to come from the
stringy moduli, it is natural to use our C-function (\ref{scc})
defined in the stringy frame to incorporate the physical effect of
dilaton. Moreover, (\ref{scc}) diverges at $H_s=0$ and is
compatible with the infinite size of the future horizon. Since the
case of $H_s=0$ is at the borderline of the accelerating Universe,
it is then reasonable to postulate the stringy C-function
({\ref{scc}) as the measure of the number of degrees of freedom of
the holographic dual theory for the accelerating universe driven
by dilaton.

     We then need to make sure the inverse C function (\ref{C}) is
monotonically decreasing such that Holographic C-theorem holds. It
turns out we need to check if
\be
\dot{H}-{d\over d-1}H \dot{\Phi}-{1\over 2}{\partial V \over
\partial \Phi}<0.
\ee
It is in general unable to determine the sign of the above
expression by using the equations of motion alone as in the case
of Einstein frame, however, one can check it explicitly for
specific potential form.

    For the case of exponential potential, the explicit time dependence
of stringy Hubble constant is
\be
H_s=g^{{-2 \over d-1}}\left(1 \pm \sqrt{d(1+\omega) \over
2}\right)H\;,
\ee
and
\be
H \propto {1\over t}\;.
\ee
where the $+$ sign is for the strong coupling branch, and the $-$
sign for the weak one.

In the case when the dilaton is driven to $-\infty$, the central
function assumes the form $C(t)=f t^{d-1}$, and $a(t)$ scales as
\be
a(t)\sim t^{{2\over d(1+\omega)}},
\label{plaw}
\ee
The exponent is greater than $1$, if the quintessence condition
$1+\omega < 2/d$ is satisfied. As $1+\omega$ approaches $2/d$, the
exponent in (\ref{plaw}) approaches $1$ and the future horizon
size defined in (\ref{hsize}) diverges as $(1/\kappa) t$, where
$\kappa=2/(d(1+\omega))-1$ approaching zero. Thus the area of the
horizon diverges according to $(1/\kappa^{d-1}) t^{d-1}$. Happily,
the central charge defined in the string frame diverges in the
exactly the same manner, namely $f\sim 1/\kappa^{d-1}$ for small
$\kappa$.

As remarked before, here $H_s$ is proportional to $H$ so that the
stringy C-theorem holds as long as $H_s>0$ and the C-theorem in
the Einstein frame holds. The latter is always true as long as
positive energy theorem is assumed. So in the strong coupling
branch, the C-theorem is true for any $\omega$ since $H_s>0$
always. However, as the string coupling becomes large one can no
longer trust the perturbative string picture, instead we expect
that an M-theory cycle will open up and it is then more natural to
use the $d+2$-dimensional dual weakly coupled theory picture to
discuss the dynamics. We will switch to this picture in the next
section. On the other hand, for the weak coupling branch the
holographic C-theorem is valid only when $H_s>0$. In this case,
the C-theorem gives no additional information as in the case of
strong coupling branch.

  We then conclude that in the case of exponential dilaton potential of
which the asymptotic geometry of the universe is not de Sitter but
Minkowski, we find an appropriate C-function defined in the
stringy frame for which the holographic C-theorem holds as long as
the stringy Hubble constant is positive. For string coupling
evolves toward the weak regime the holographic picture defined
above is valid only for the accelerating universes in the Einstein
frame.

In the above discussion we have assumed the existence of an
attractive fixed point in the solution space such that the ratio
$p/\rho$(or $\rho/V$) is constant and the above check for the
holographic C-theorem is valid. However, this is not always true
if there is no fixed point such that there is a long transient for
$\rho/V \rightarrow constant$. A special case is for constant
potential $V$ so that one can solve
\be
\dot{\Phi}\propto e^{-d\lambda}
\ee
such that $\rho/V=1+\dot{\Phi}^2/V(d-1)$ is not a constant. In
this case, for the strong coupling branch, the holographic
C-theorem still holds and the stringy Hubble constant is positive
definite.  On the other hand, for the weak coupling branch, the
stringy Hubble constant
\be
H_s \propto \sqrt{\rho \over d(d-1)}-\sqrt{(d-1)(\rho-V)}\;,
\ee
which can not be always positive definite and can easily become
negative in the early universe where $\rho>>V$, however, the
string coupling is strong then and the perturbative picture breaks
down. To check the C-theorem we find that
\be
\dot{H}-{d\over d-1}H \dot{\Phi}={-\sqrt{\rho-V} \over
d-1}(\sqrt{\rho-V}-\sqrt{d\rho})\;,
\ee
which is positive in general and the holographic C-theorem is
violated. Note that in the late time the above quantity approach
to zero as $\rho \rightarrow V$, which signals a finite central
charge for the UV fixed point, the dual CFT of the de Sitter
space. One can also see this in the Einstein frame.

  Another example without the attractive fixed point in the solution
space is the tracker solution associated with the power-law
potential discussed in the last section. The corresponding Hubble
constant in the stringy frame is
\be
H_s^{(T)}={ e^{-Z/\sqrt{d-1}}\over \sqrt{d-1}}(\sqrt{\rho_0 \over
d}(-Z)^{-\delta/2}-\sqrt {C_0} (-Z)^{-(\delta+2)/2})\;.
\ee
Obviously when the 2nd term dominates in the early universe
$H_s^{T}$ becomes negative, however, then the perturbative picture
breaks down. On the other hand, in the late time $H_s^{(T)}$ is
positive and the the perturbative picture is good. Similarly, for
the C-theorem to hold it is easy to see that one should require
\be
{-1\over d-1}(-Z)^{-\gamma}+{\gamma\over
2\sqrt{d-1}}(-Z)^{-(\gamma+1)}<0\;.
\ee
Again, the inequality is easily violated in the early Universe but
holds true in the late time.

  Besides the validity of the perturbative string picture in the bulk,
the holographic RG flow and the C-function are highly sensitive to
the asymptotical geometry which classifies the bulk geometry and
the dual CFT.  Even though the C-theorem seems being violated in
the early universe for the tracker potential, it holds at the late
time where the dark energy dominates and the asymptotical geometry
is the same as the one for the exponential-law potential.
Therefore, the exponential and the tracker cases should share the
same holographic picture in the far future. This justifies that
the choice of the stringy C-function is also good for the tracker
case.

\section{Strong Coupling Branch from Its Dual M-theory}
As remarked in the last section for some case the string coupling
will grow as the Universe evolves so that the perturbative string
picture can not be relied on any more. If so, the non-perturbative
string modes become light and a M-theory cycle opens up so that
the effective theory becomes a weakly coupled (d+2)-dimensional
Einstein gravity
\be
S_M=-\int d^{d+2}x
\sqrt{-\tilde{G}}[\tilde{R}-\sum^D_{i=1}(\tilde{\nabla}\sigma_i)^2
+\tilde{V}(\sigma_i)]\;,
\ee
where the tilde quantities are with respect to the
(d+2)-dimensional Einstein metric $\tilde{G}_{ab}$. After
compactifying on the following metric ansatz
\be
\tilde{G}_{ab}dx^adx^b=e^{{2\over \sqrt{d}}\Phi}dy^2+e^{{-2\over
d-1} (1+{1\over \sqrt{d}})\Phi}
\hat{g}_{\mu\nu}dx^{\mu}dx^{\nu}\;,
\ee
the action $S_M$ will reduce to the stringy action
(\ref{stringya}) with the potentials are related by
$U(\Phi,\sigma_i)=e^{{-2\over d-1}(1+{1\over
\sqrt{d}})\Phi}\tilde{V}(\sigma_i)$ at the classical level, which
leads to the relation between the potential $V$ in the Einstein
frame and $\tilde{V}$ as $V(\Phi,\sigma_i)=e^{{-2\over
(d-1)\sqrt{d}}\Phi}\tilde{V}(\sigma_i)$. This relation is too
restrictive and could be modified by the quantum correction, also
by the mixing of the dilaton and the other internal moduli, we
will assume the complications in the following discussions such
that the potential $V$ is a generic one.

Note that the metric component $\tilde{G}_{yy}$ is increasing as
$\Phi$ grows, which reflects the open-up of the M-theory cycle as
string coupling becomes strong. Then the Hubble constant and the
Newton constant in the M-frame with respect to the metric
$\tilde{G}_{\mu\nu}= e^{{-2\over d-1}(1+{1\over
\sqrt{d}})\Phi}\hat{g}_{\mu\nu}=e^{{-2\over(d-1)\sqrt{d}}\Phi}
g_{\mu\nu}$ in terms of the quantities defined in the Einstein
frame with respect to the metric $g_{\mu\nu}$ are
\bea
\label{HM}
H_M&=&g^{{-1 \over (d-1)\sqrt{d}}}(H-{1\over
(d-1)\sqrt{d}}\;\dot{\Phi})\;,
\\
G_M&=&g^{{-1\over \sqrt{d}}}G^{'}_N\;,
\eea
where $G^{'}_N$ is the (d+2)-dimensional Newton constant without
moduli dependence.

The inverse C-function in the M-frame is
\be
H_M^{d-1}G_M=G^{'}_N(H-{1\over
(d-1)\sqrt{d}}\;\dot{\Phi})^{d-1}\;.
\ee
Similar to the case in stringy frame, the inverse C-function in
M-frame is again of no $g$ factor despite of the $g$ factor in
$H_M$. As remarked before, this is a good sign for the choice of
the C-function, otherwise, the complication of the $g$ factors
will make difference between the leading scale behaviors of the
C-functions defined in the M-frame and in the Einstein frame.
Unlike the stringy frame case, we do not have the test of the
infinite future horizon for the M-frame since $H_M>0$ always as
shown below.

Although there is a relative sign in (\ref{HM}) it is
straightforward to see that $H_M\ge0$ where the equality holds
only for $V=0$. This means that the universe in the M-frame is
always expanding. Explicitly, by choosing positive $\dot{\Phi}$ we
get
\be
H_M=g^{{-1 \over (d-1)\sqrt{d}}}\left(1- \sqrt{1+\omega \over
2}\right)H\;,
\ee
so that $H_M>0$ as long as $\omega<1$. Note that $\omega$ is
always smaller than or equal to one for non-ghost canonical scalar
even it is a function of time. So, the fact of $H_M>0$ is true
also for non-exponential potential. Similarly to the discussion in
the stringy frame, the C-theorem in the M-frame holds true always
because $H_M>0$ and $H_M \propto H$.

\section{Conclusion}

We know little about nonsupersymmetric string theory with a small,
positive vacuum energy. There are two lines one may follow to gain
more knowledge. One is to follow basic rules available in string
theory to study supersymmetry breaking to get as realistic as
possible. Another is to follow what cosmology has to teach us, to
push the dichotomy faced in string cosmology as far as possible.
The latter line is more ``phenomenological". Approaches to
formulating a holographic dual of an accelerating universe are
more in this spirit. Even in such a ``phenomenological" approach,
one has very few handles. It appears then that the C-function and
the corresponding RG flow is one of such handles.

We explored alternative C-functions in a general situation when
the fate of the universe is not exactly de Sitter. The C-function
in a universe driven to the weak coupling regime sounds more
attractive, since the C-theorem results in a genuine constraint on
possible potentials for the dilaton. It is certainly desirable to
consider even more general situations when there are more than one
moduli.

In a four dimensional universe, we do not have a first principle
based upon which a general C-function can be constructed. As a
working hypothesis, a general C-function ansatz contains only
metric and scalar fields and their first derivatives. The first
derivatives can be replaced by their corresponding beta functions
in the RG language. Then the C-function in turn determines the
beta functions in the following fashion
$$\beta^i=G^{ij}{\partial C\over \partial g^j},$$
where $g^i$ stand for metric and scalar fields. However, without knowing
$G^{ij}$ it is impossible to deduce the C-function by comparing the above
gradient flow equations and equations of motion derived from a classical
action.

For the above reason, it is then desirable to make reasonable
guesses in various situations and to check the consequences of
these guesses, in order to make progress. The work presented in
this note may be viewed as a step in this direction.

\bigskip
\noindent{\bf Acknowledgments} ML was supported by a grant of NSC,
and by a ``Hundred People Project'' grant of Academia Sinica and
an outstanding young investigator award of NSF of China. FLL was
supported by the NSC grant No. NSC89-2112-M-032-002. The authors
would like to thank the courtesy of the CTP at NTU, the NCTS, and
the CosPA project, Taiwan. \vfill\eject

\end{document}